\documentclass[12pt]{article}
\usepackage[]{graphicx}
\usepackage[]{color}

\usepackage{amsmath}
\usepackage{graphicx,psfrag,epsf}
\usepackage{enumerate}
\usepackage{natbib}
\usepackage{url} 

\newcommand{\blind}{0}

\makeatletter
\def\maxwidth{ %
  \ifdim\Gin@nat@width>\linewidth
    \linewidth
  \else
    \Gin@nat@width
  \fi
}
\makeatother

\definecolor{fgcolor}{rgb}{0.345, 0.345, 0.345}

\usepackage{framed}
\makeatletter
\newenvironment{kframe}{%
 \def\at@end@of@kframe{}%
 \ifinner\ifhmode%
  \def\at@end@of@kframe{\end{minipage}}%
  \begin{minipage}{\columnwidth}%
 \fi\fi%
 \def\FrameCommand##1{\hskip\@totalleftmargin \hskip-\fboxsep
 \colorbox{shadecolor}{##1}\hskip-\fboxsep
     \hskip-\linewidth \hskip-\@totalleftmargin \hskip\columnwidth}%
 \MakeFramed {\advance\hsize-\width
   \@totalleftmargin\z@ \linewidth\hsize
   \@setminipage}}%
 {\par\unskip\endMakeFramed%
 \at@end@of@kframe}
\makeatother

\definecolor{shadecolor}{rgb}{.97, .97, .97}
\definecolor{messagecolor}{rgb}{0, 0, 0}
\definecolor{warningcolor}{rgb}{1, 0, 1}
\definecolor{errorcolor}{rgb}{1, 0, 0}
\newenvironment{knitrout}{}{} 

\usepackage{alltt} 

\usepackage{natbib} 
\usepackage[colorlinks=true, citecolor=blue, linkcolor=blue]{hyperref}

\usepackage{booktabs}
\usepackage{array}
\newcolumntype{M}{>{\centering\arraybackslash}m{\dimexpr.05\linewidth-2\tabcolsep}}

\newcolumntype{C}[1]{>{\centering}m{#1}}

\usepackage{graphicx}
\DeclareGraphicsExtensions{.eps, .pdf}
\graphicspath{{figures/}}

\usepackage{wrapfig,float}
\usepackage{caption}
\usepackage{subcaption}

\usepackage{todonotes}

\usepackage[margin = 1in]{geometry}

\usepackage{setspace}
\doublespace

\usepackage{bm}
\usepackage{amstext}
\usepackage{amssymb}
\usepackage{amsmath}
\usepackage{amsfonts}
\usepackage{multirow}

\IfFileExists{upquote.sty}{\usepackage{upquote}}{}
\begin{document}

\def\spacingset#1{\renewcommand{\baselinestretch}%
{#1}\small\normalsize} \spacingset{1}

\if0\blind
{
  \title{\bf Variations of Q-Q Plots -- The Power of our Eyes!}
  \author{{Adam Loy, Lendie Follett, Heike Hofmann}\thanks{
    Adam Loy is an Assistant Professor in the Department of Mathematics, Lawrence University, Appleton, WI, 54911 (e-mail: adam.m.loy@lawrence.edu);  Lendie Follett is a Ph.D. student in the Department of Statistics and Statistical Laboratory, Iowa State University, Ames, IA 50011-1210; Heike Hofmann is a Professor in the Department of Statistics and Statistical Laboratory, Iowa State University, Ames, IA 50011-1210. This work was funded in part by National Science Foundation grant DMS 1007697. All data  in the study was collected  with approval from the internal review board IRB 10-347.}\hspace{.2cm}\\
}
\date{September 8, 2014}
  \maketitle
} \fi

\if1\blind
{
  \bigskip
  \bigskip
  \bigskip
  \begin{center}
    {\LARGE\bf Variations of Q-Q Plots -- the Power of our Eyes!}
\end{center}
  \medskip
} \fi

\bigskip
\begin{abstract}
In statistical modeling we strive to specify models that resemble data collected in studies or observed from processes. Consequently, distributional specification and parameter estimation are central to parametric models. Graphical procedures, such as the quantile-quantile (Q-Q) plot, are arguably the most widely used method of distributional assessment, though critics find their interpretation to be overly subjective. Formal goodness-of-fit tests are available and are quite powerful, but only indicate whether there is a lack of fit, not why there is lack of fit. In this paper we explore the use of the lineup protocol to inject rigor to graphical distributional assessment and compare its power to that of formal distributional tests. We find that lineups of standard Q-Q plots are more powerful than lineups of de-trended Q-Q plots and that lineup tests are more powerful than traditional tests of normality. While, we focus on diagnosing non-normality, our approach is general and can be directly extended to the assessment of other distributions.
\end{abstract}

\vfill

\clearpage
\spacingset{1.45}

\section{Introduction}

In statistical modeling we strive to specify models that resemble data collected in studies or observed from processes. Consequently, 
distributional specification and parameter estimation are central to parametric models.
The statistical modeling process is cyclical \citep{tukey:eda}, so after parameters are estimated and the model is checked, the process might continue through another cycle with a refined model formulation. Model checking is central to statistical modeling; in particular, any conclusions based on a model depend  on  correct distributional specifications. For example, prediction intervals in the classical regression setting depend directly on the assumption of normality, so they are quite sensitive to departures from normality. 

Graphical procedures, such as the quantile-quantile (Q-Q) plot \citep{Wilk:1968}, are arguably the most widely used method of distributional assessment, though critics find their interpretation to be overly subjective. Formal goodness-of-fit tests are available and are quite powerful, but only indicate whether there is a lack of fit, not why there is lack of fit. For example, the Shapiro-Wilk test \citep{Shapiro:1965kt} is a powerful test of normality, but does not indicate what feature of the distribution is non-normal, so a plot, such as a Q-Q plot, must be rendered after any rejection. 

In this paper we explore the use of the lineup protocol \citep{buja:2009hp} to inject rigor to graphical distributional assessment and compare its power to that of formal distributional tests. We focus on diagnosing non-normality, so our discussion centers around the normal Q-Q plot, but our approach is general enough and can be directly extended to the assessment of other distributions.

We will first discuss  tests for normality, both from a numerical and graphical viewpoint, and then formally introduce the lineup protocol in the setting of quantile-quantile plots used for this paper.



\subsection{Classical tests of normality}
Numerous tests have been proposed to test whether a random sample comes from a normal distribution. In this section we review commonly used tests of normality. 

A series of distributional tests focuses on the difference between the empirical and theoretical distribution functions. More formally, let $F_n$ be the empirical distribution function (ECDF) based on a sample size of $n$, and $F$ be the hypothesized/true distribution. The absolute difference between the two distribution functions for each sample point, $\left| F_n(x_i) - F(x_i) \right|$, is the main contributor for the test statistics of the Kolmogorov-Smirnov \cite[KS-test,][]{kolmogorov:1933, smirnov:1948}, the Lilliefors \cite[LF-test, ][]{lilliefors}, the Anderson-Darling \citep[AD-test,][]{adtest:1954}, and the Cram\'{e}r-von-Mises tests \citep[CVM-test,][]{cramer:1928, mises:1928}, as shown in table~\ref{tab:tests}.

The KS test uses the maximal  difference, 
regardless of the range of the sample---i.e. a difference, $D$, observed in either tail of the distribution carries the same weight and is interpreted in the same way as a difference, $D$, in the center of the distribution. While the KS test allows for the adjustment of the parameters of the normal distribution to the sample mean and variance, it is more appropriate to use the LF test for this purpose. LF and KS share the same test statistic, but the sampling distribution in the LF test statistic is adjusted for the two additional parameters.  AD and CVM  are both based on the total area between the hypothesized distribution function and the empirical distribution function. Compared to the KS  test,  the CVM test downplays the effects in the tails of a (normal) distribution, while the AD test upregulates the tail effect using a weighting of $1/\left(F(x)(1 - F(x)\right)$ across the range of the sample.

\begin{table}
\centering
\caption{\label{tab:tests} Four prominent tests for normality based on the difference between empirical and hypothesized distribution function. An overview of the performance and power of these tests can be found in \citet{stephens:1974}.}
\begin{tabular}{lrl}\hline
Test && Statistic\\\hline\hline
Kolmogorov-Smirnov & $D =$ & $ \sup_{1 \le i \le n} \left | F_n(x_i) - F(x_i)\right|$ \\
Lilliefors & $D =$ & $ \sup_{1 \le i \le n} \left | F_n(x_i) - F(x_i)\right|$ \\
Anderson-Darling & $A =$ & $ n \int_{-\infty}^{+\infty} \left | F_n(x) - F(x)\right|^2/\left(F(x)(1 - F(x)\right) dF(x)$\\
Cram\'{e}r-von-Mises & $C =$ & $n \int_{-\infty}^{+\infty} \left | F_n(x) - F(x)\right|^2 dF(x)$ \\\hline
\end{tabular}
\end{table}
%

The Shapiro-Wilk test \cite[SW-test,][]{Shapiro:1965kt} does not utilize deviations from the theoretical distribution function, rather it focuses on the linearity of a normal Q-Q plot. Under normality, a set of observations, $x_1, \ldots, x_n$, can expressed as $x_i = \mu + \sigma z_i$, where $z_i$ is a quantile from the standard normal distribution. The Shapiro-Wilk test compares (up to a constant of proportionality, $c$) two estimates for $\sigma$: the best linear unbiased estimate obtained from a generalized least squares regression of the sample order statistics on their expected values, denoted $\widehat{\sigma}$, and the sample standard deviation, $s$.
\[
  W = \frac{(c \widehat{\sigma})^2}{s^2} = \frac{b^2}{s^2}
\]
For a sample drawn from a normal distribution, $b^2$ and $s^2$ are, up to a constant, estimating the same quantity, whereas the two estimators will generally not be estimating the same quantity under non-normality. The SW test has been shown to be the most powerful in assessing non-normality \citep{stephens:1974, razali:2011}.

In Section~\ref{sec:power2} we will return to these tests in order to assess the effectiveness of different variations of standard Q-Q plots.

\subsection{Q-Q Plots}

Standard quantile-quantile (Q-Q) plots \citep{Wilk:1968} are an essential tool for  visually evaluating a specific distributional assumption.  A Q-Q plot  is constructed from a sample, $x_1, \ldots, x_n$, by plotting the theoretical quantiles, $F^{-1}(F_n(x_i))$, against the sample quantiles, $x_{(i)}$. If the empirical distribution, $F_n$, is consistent with the theoretical distribution, $F$, the points in the Q-Q plot fall on the line of identity. 
For any sample tested against a distribution within a location-scale family, such as a normal, log normal, or exponential distribution, the sample quantiles still fall on a line when plotted against the theoretical quantiles of any of the family's member distributions. Plotting the empiricial quantiles of a normally distributed sample $x \sim N(\mu, \sigma^2)$ against the quantiles of a standard normal will result in a line, where  the slope is an estimate of $\sigma$, and the intercept estimates $\mu$. Visually  the only change in the Q-Q plot is a  change in the scale of the $y$-axis. We can therefore employ Q-Q plots in the more general framework of testing the distribution of a sample for normality similar to standard normality tests, such as the AD, LF, CVM, and SW tests. We do have to make a decision with respect to the exact parameters of the normal distribution we test against when we plot a line alongside the points in the Q-Q plot for additional comparison purposes, i.e.~the parameters $\mu$ and $\sigma$ have to be estimated from the sample. In Q-Q plots, variability is based on a robust measure of spread given as the ratio of the inter-quartile ranges (IQRs) of the empirical and theoretical distributions: $\left(F^{-1}_n(0.75) - F^{-1}_n(0.25)\right) / \left(F^{-1}(0.75) - F^{-1}(0.25)\right)$ \citep{becker:s}. 

Based on a visual inspection in a Q-Q plot, a
 sample is therefore considered to be consistent with a normal distribution if the empirical and theoretical quantiles fall close to the line representing the theoretical distribution.  This decision is helped additionally by an assessment of
 whether the points fall inside the envelope of 95\%  pointwise confidence intervals \citep[][p.~150--154]{Davison:1997}.

In assessing differences between points and lines, onlookers have a tendency to evaluate the shortest, i.e.~orthogonal, distance, even when asked to evaluate differences based on vertical distance \citep{sineillusion, robbins:2005, cleveland:1984}. 
In so-called {\it de-trended Q-Q plots} \citep[][p.~25--26]{thode:2002} the $y$ axis is changed to show the difference between theoretical and sample quantiles. The line of the theoretical distribution therefore falls onto the $x$ axis, and (vertical) differences between empirical and theoretical distribution coincide with the orthogonal distance. 
De-trending should aid in the visual assessment between the empirical CDF and theoretical CDF. This also follows the general standard graphical recommendation to directly plot the aspect of the data we want to show rather than asking audiences to derive it \citep{wainer:2000}.  Another point in favor of this design is that it makes better use of the available space.

\begin{figure}
\centering
\includegraphics[width=0.3\textwidth]{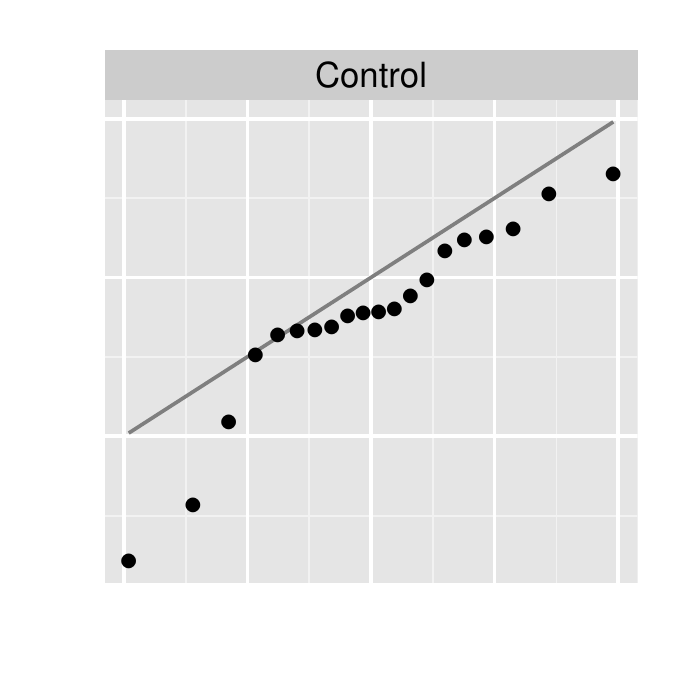}
\includegraphics[width=0.3\textwidth]{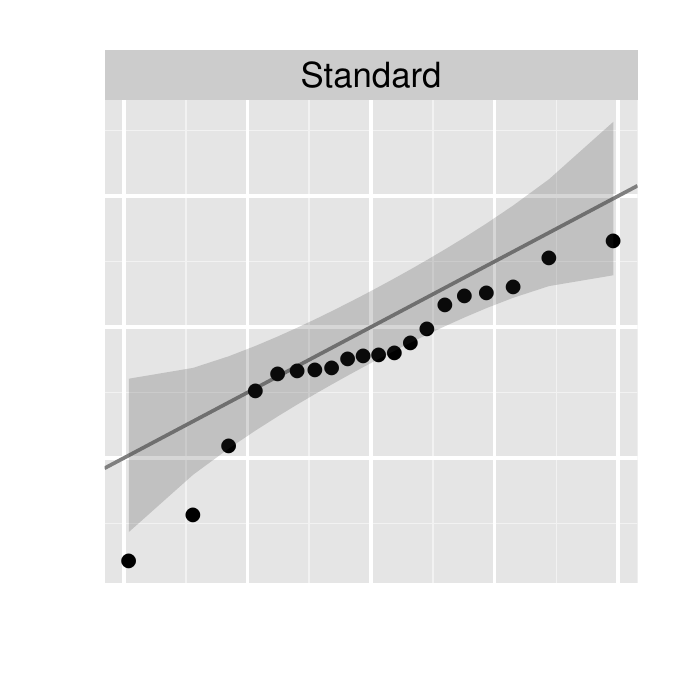}
\includegraphics[width=0.3\textwidth]{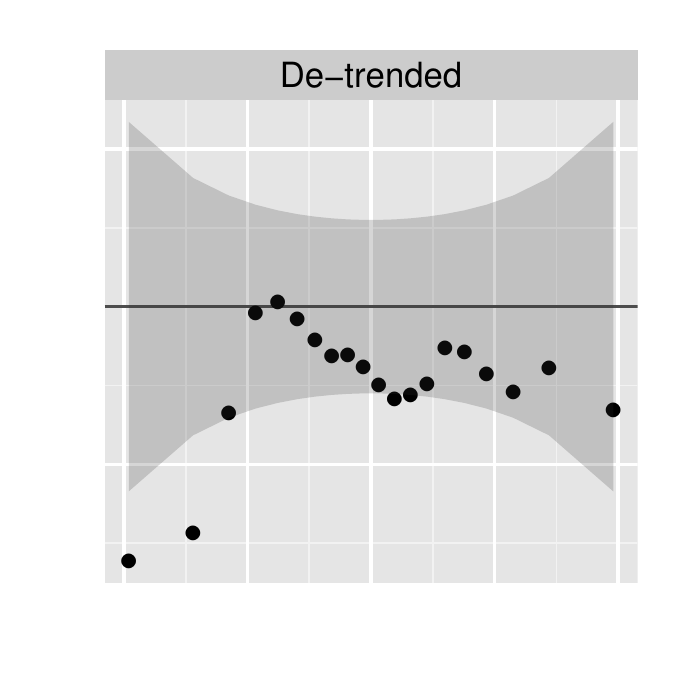}
\caption{ \label{qqplots} Three versions of Q-Q plots: control, standard, and de-trended.}
\end{figure}

In this paper we investigate the effectiveness and power of the modifications made to Q-Q plots.
Examples of the three versions of Q-Q plots under consideration are displayed in Figure~\ref{qqplots}, and include (from left to right): a \emph{control} Q-Q plot, a \emph{standard} Q-Q plot with an added grey band representing a 95\% pointwise confidence region \citep{Davison:1997}
based on the estimated standard error of the order statistics for an independent sample from the theoretical distribution, and a \emph{de-trended} Q-Q plot. Note that all Q-Q plots in Figure~\ref{qqplots} are constructed from the same data. 

In order to objectively evaluate  the three designs and quantify their effectiveness we make use of {\it lineup tests}.

\subsection{Lineup Tests}
Lineup tests have been introduced by \citet{buja:2009hp} to evaluate and quantify the significance of graphical findings. The idea behind a lineup test is that of a police lineup: the chart of the observed data is placed randomly among a set of so-called \emph{null charts}, showing data created consistently with the null hypothesis. In the setting of a lineup of normal Q-Q plots, the null hypothesis  is either that $F$ is standard normal or that $F$ is normal with parameters based on sample mean and variance.
If the `suspect'---i.e. the plot of the observed data---can be identified from the null charts, this counts as evidence against the null hypothesis. Multiple identifications of the data by independent observers then lead to a rejection of the null hypothesis. 
The lineup protocol also allows for an assessment of the power of a lineup \citep{mahbub:2013},  
and by showing different renderings of the exact same data in lineups we can evaluate the power  of different designs \citep{Hofmann:2012ts}.

In considering the power of a lineup, we need to estimate the probability, $p_i$, that observer $i$ identifies the data from the lineup. If the observer is just guessing, this probability is $1/m$, where $m$ is the number of plots in the lineup.
The power of a lineup is then given as the probability to reject the null hypothesis. Let $Y$ be the number of identifications of the data plot in $N$ independent evaluations, and let $Y \sim F_N$. The power of the lineup is then the probability that more than $y_\alpha$ out of $N$ observers
choose the true plot, or more formally
\begin{equation}\label{eqn:power}
\widehat{\text{Power}} = \text{Power}_{N} = 1 - F_{Y} (y_{\alpha}),
\end{equation}
where $y_\alpha$ is the critical value for a given significance level $\alpha$, i.e.~$P(Y >  y_{\alpha}) \le \alpha$. $Y$ is composed of the sum of $N$ observers' (binary) decisions $Y_i \sim B_{1, p_i}$, where  $p_i$ is the probability that individual $i$ chooses the data plot. This probability  depends both on the strength of the signal in the data plot and an individual's visual ability.
Assessing this ability requires that each individual evaluates multiple lineups. 
If that is not possible, we must assume that all participants share the same ability, $p$. 
Similar to classical inference, we can make use of power to assess the sensitivity of tests. This allows us to make decisions about designs for particular tasks by evaluating lineups displaying  the same data in different types of displays \citep{Hofmann:2012ts}. 

In the next section, we describe the simulation study used to compare the three Q-Q plot designs. An initial comparison of the three designs is also given. We use a generalized linear mixed model to compare the power of the three designs in section~\ref{sec:power1}, and also explore the feedback of the independent observers to compare the rationale for plot selection (i.e., rejection of normality). Finally, we compare the power of a lineup test of normality to the classical normality tests in section~\ref{sec:power2}, and outline areas for future research in section~\ref{sec:discussion}.


\section{Simulation Setup and Model}\label{sec:simu}

To further develop the assessment of normality using lineups, we conducted a study comparing the three different versions of the Q-Q plot.

To investigate the power of the three different Q-Q plot versions, we sampled data from a $t$ distribution with varying degrees of freedom and sample sizes, and included a Q-Q plot of these data in a lineup of null charts drawn from standard normal samples of the same size.
For lineup tests it is of extreme importance to consider the generation of the null sets and the construction of the plots in the lineup. 
Null data is created conistently with the null hypothesis. Here, we have two different null hypotheses to consider:

\begin{itemize}
\item{\bf Situation I:}
$H_0: F = N(0,1)$  \\
Null samples are drawn from a standard normal distribution; the reference line is the line of identity. Lines and envelopes are the same across all panels, in particular, all panels have the same scale. 
\item{\bf Situation II:} $H_0: F = N(0,S^2)$ \\
$S$ is based on the interquartile range of the data; null samples are drawn from $N(0, S^2)$. \\
The reference line has a slope of $S$ (and an intercept of 0).  All panels have the same scale. 
\end{itemize}

\begin{figure}[hbt]

\begin{subfigure}{0.5\textwidth}
\begin{knitrout}
\definecolor{shadecolor}{rgb}{0.969, 0.969, 0.969}\color{fgcolor}
\includegraphics[width=\maxwidth]{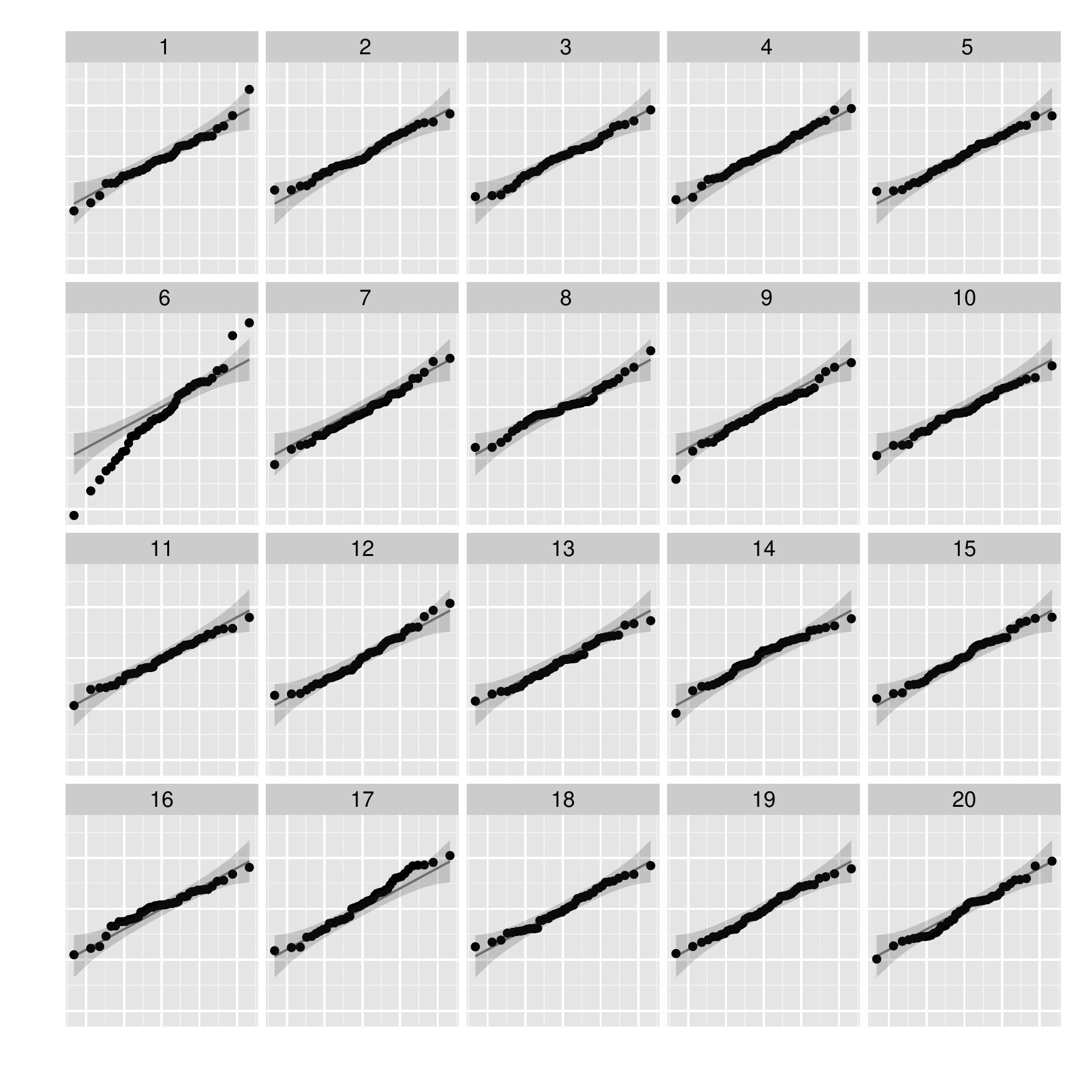} 

\end{knitrout}
\end{subfigure}
\begin{subfigure}{0.5\textwidth}
\begin{knitrout}
\definecolor{shadecolor}{rgb}{0.969, 0.969, 0.969}\color{fgcolor}
\includegraphics[width=\maxwidth]{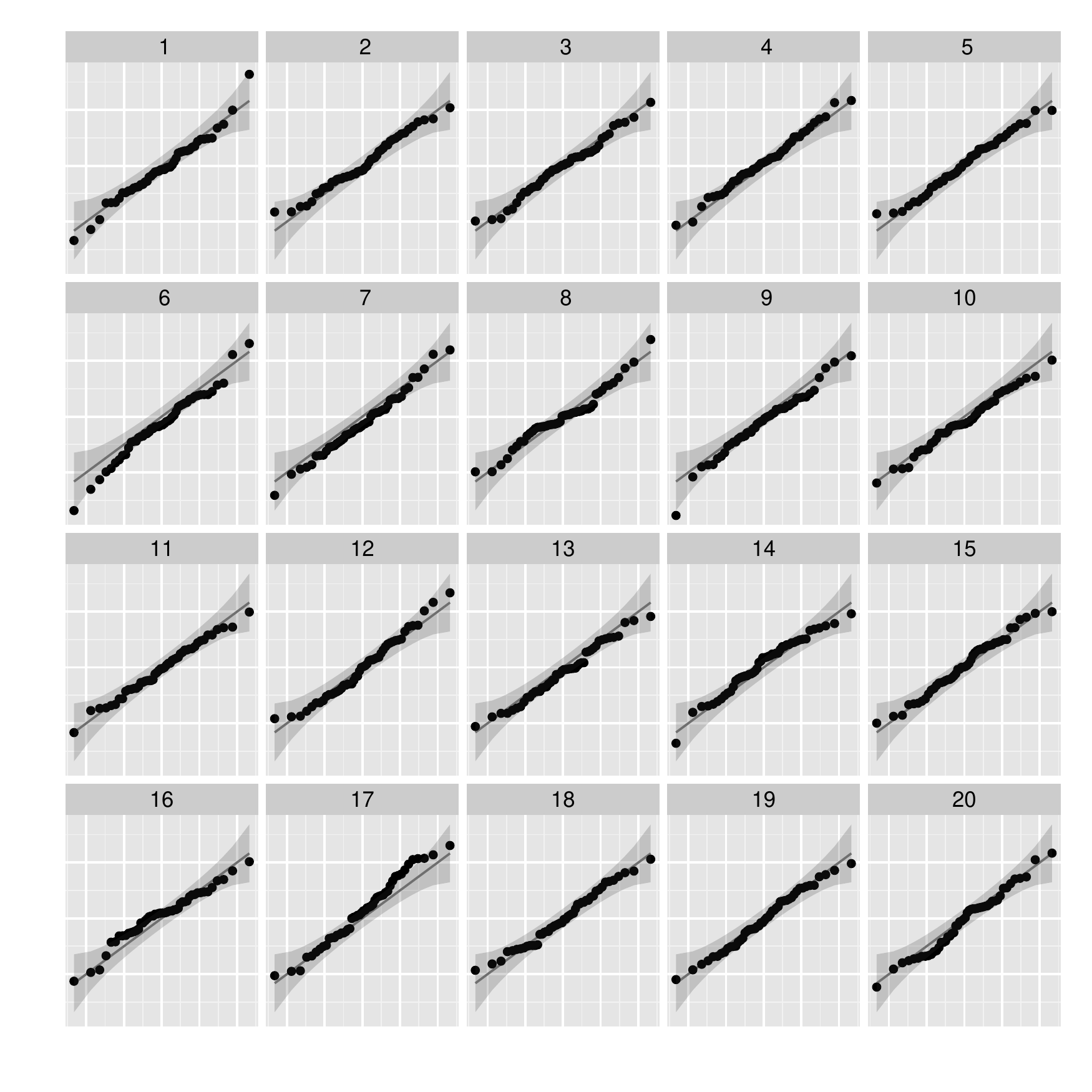} 

\end{knitrout}
\end{subfigure}
\caption{\label{fig:lps} Lineup plots of standard Q-Q plots. The observed data is the same, but  reference lines and envelopes are based on a standard normal distribution on the left; while  reference lines and envelopes for the lineup on the right are based on a normal distribution $N(0, \widehat{S}^2)$, where $\widehat{S}$ is based on the IQR of the observed data.
The observed data in both lineups is displayed in panel \#$(3^2 - 3)$. }
\end{figure}

Examples for both hypotheses are shown in figure~\ref{fig:lps}. Both lineups show the same dataset (in panel \#$(3^2-3)$). On the left the data stand out (all 33 observers picked the data plot), i.e.~we reject the null hypothesis of a standard normal distribution. On the right, the data do not stand out (only 3 out of 27 observers picked the data); thus, we do not reject the hypothesis of a normal distribution with parameters $\mu=0$ and $\widehat{\sigma}=1.578$. 

Note that the above list of hypotheses is not exhaustive. Any theoretical distribution in Q-Q plots corresponds to a  hypothesis test against that distribution. As long as there is a method to generate samples under the null hypothesis, we do not even need to know the exact distribution. This allows us to assess situations in which we only have approximate or asymptotic results, which are hard, if not impossible, to investigate with the (small) finite samples we typically deal with in practice.

Note that IQR is used here in estimating scale---this is standard practice for Q-Q plot. Robust estimation of the variance is preferred for better assessment of the tails and outliers of the empirical distribution. We could use alternative estimators for variance, such as median absolute deviation (MAD) or adjusted MAD \citep{rousseeuw}, but this will likely also change the power of the corresponding lineup.

Next, we model the aforementioned probability $p_i$ with which observer $i$ picks the true data from a lineup. 
Let $X_i \sim B_{1, \pi_i}, 1 \le i \le n$, where $X_i$ is the binary decision on the $i$th evaluation and $\pi_i$ is the probability with which the observer chooses the data plot. This probability is influenced by a number of factors:

\begin{center}
\begin{tabular}{lp{5in}}
$\tau$ & the design used in the lineup (Control, Standard, De-trended), \\
&  the specific parameters under which the data for the lineup were created: \\
&  $\delta$ \ \ \ degrees of freedom (2, 5, 10) {of the $t$ distribution} and \\
&  $\nu$  \ \ \ sample size (20, 30, 50, 75), \\
$d$ &  the level of difficulty based on the actual sample, and \\
$u$ & the users' subjective abilities.
 \end{tabular}
\end{center}
The combination of different levels of sample size and degrees of freedom of the $t$ distribution result in 12 parameter settings. Under each setting, we  generated data twice. Additionally, we made use of two different sets of null data for each sample, yielding 48 different sets of data, which we render in each of the variations, resulting in 144 different lineups.

Using  Amazon MTurk \citep{amazon}, 674 independent observers were recruited and asked to evaluate ten lineups each. 
Half of the lineups that observers were shown allowed multiple choices of plots from a lineup for the final answer. While most participants still chose only a single plot, in the analysis we dealt with multiple answers to a lineup by using a weighting variable defined as the reciprocal of the number of answers given by a participant.

\begin{knitrout}
\definecolor{shadecolor}{rgb}{0.969, 0.969, 0.969}\color{fgcolor}\begin{kframe}
\begin{verbatim}
## Loading C code of R package 'Rmpfr': GMP using 64 bits per limb
\end{verbatim}
\end{kframe}
\end{knitrout}

\begin{figure}
\centering
\begin{subfigure}[b]{.3\textwidth}
\begin{knitrout}
\definecolor{shadecolor}{rgb}{0.969, 0.969, 0.969}\color{fgcolor}
\includegraphics[width=\maxwidth]{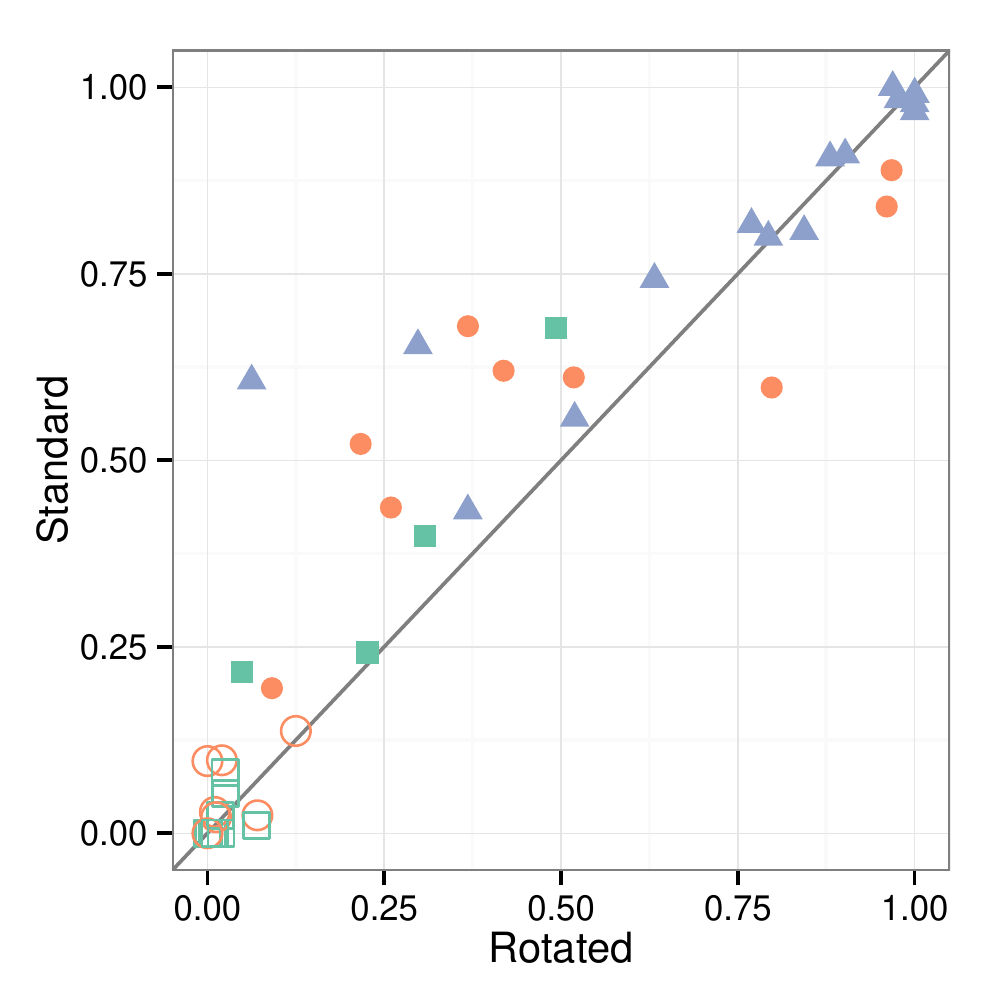} 

\end{knitrout}
\end{subfigure}
\begin{subfigure}[b]{.3\textwidth}
\begin{knitrout}
\definecolor{shadecolor}{rgb}{0.969, 0.969, 0.969}\color{fgcolor}
\includegraphics[width=\maxwidth]{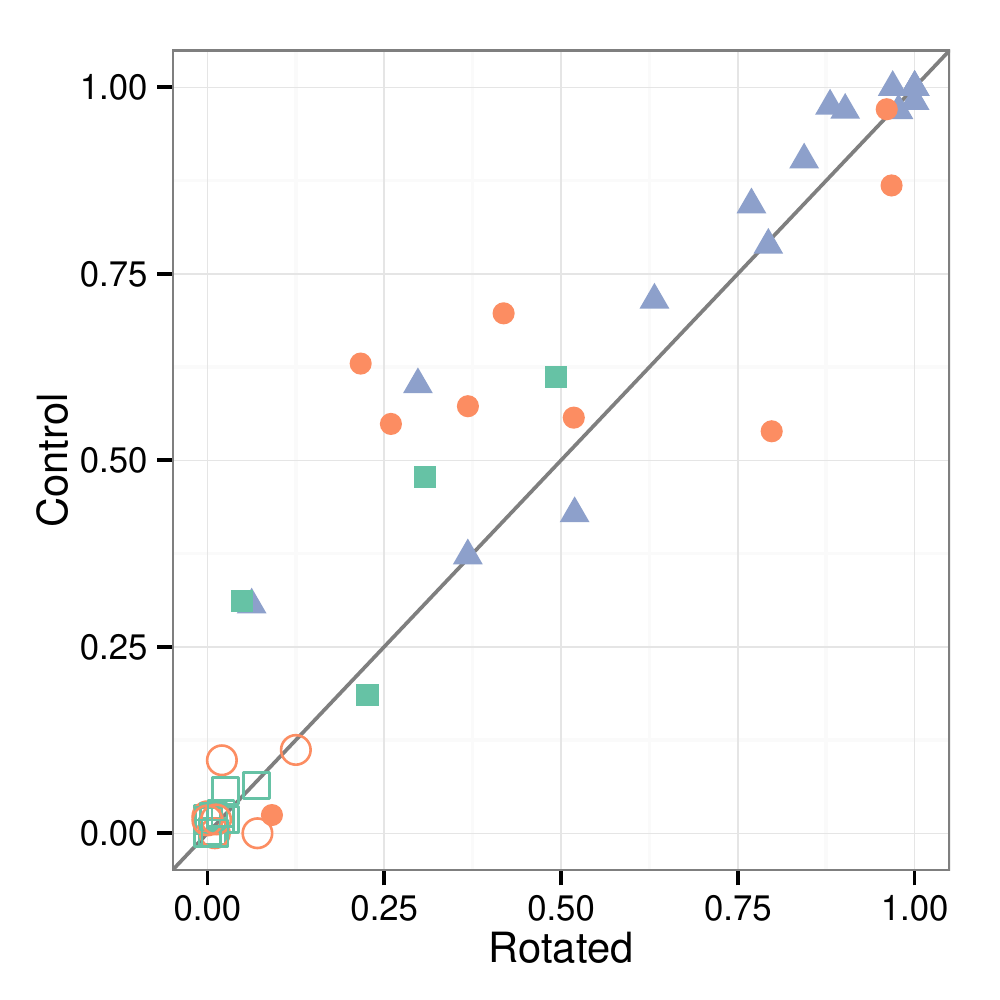} 

\end{knitrout}
\end{subfigure}
\begin{subfigure}[b]{.3\textwidth}
\begin{knitrout}
\definecolor{shadecolor}{rgb}{0.969, 0.969, 0.969}\color{fgcolor}
\includegraphics[width=\maxwidth]{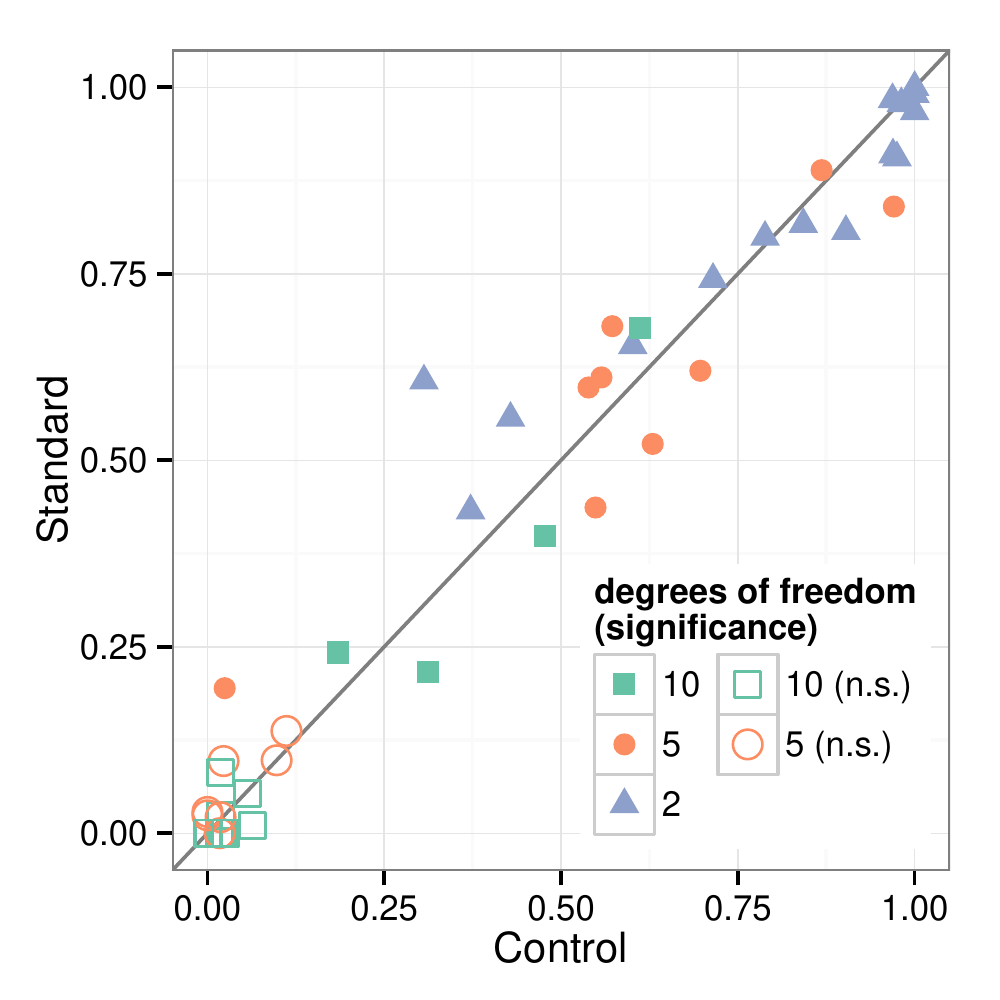} 

\end{knitrout}
\end{subfigure}
\caption{\label{fig:compare}Proportions of successful evaluation of the same data in the three different variations of Q-Q plots. Standard and control displays exhibit the highest correlation. De-trended Q-Q plots agree with decisions made based on Q-Q plots in the control or standard design, but display lower rates of correct responses in the `middle' field. Significances are based on lineup evaluations in the standard design. }
\end{figure}

Figure~\ref{fig:compare} shows proportions of correct evaluations of the lineups under the three different variations of Q-Q plots. All three versions provide highly correlated results, and largely agree for extreme decisions (all correct/all wrong evaluations). In the middle range de-trended Q-Q plots perform  worse than either standard or control Q-Q plots. Lineups of data samples from a $t_2$-distribution  are all rejected in lineups under the standard design, while most of the $t_{10}$ samples go undetected. Lineups showing a sample from a $t_5$ distribution cover the whole range.
We base an evaluation of the three different Q-Q plot designs on the premise that if participants find it easier to identify the data plot under one lineup over another lineup (given identical data underlying the lineups), the first lineup uses the better design.

Let $Y_i$ be the outcome of the $i$th evaluation. Then $Y_i$ is a Bernoulli variable, where $\pi_i$ denotes the probablity of identifying the data plot from the lineup; i.e.~$P(Y_i = 1) = \pi_i = E[Y_i]$.   
The probability of identifying the lineup is affected by several factors: (a) the strength of the signal, i.e.~degrees of freedom of the $t$ distribution, and the sample size, (b) a human factor, i.e~the visual ability of the observer, and (c)  the `lineup factor': depending on which $m-1$ representatives of the test distribution the null plots show, lineups of the same data plot can have different difficulty. We capture all of this in a logistic regression model with fixed effects for signal strength and random effects for lineup difficulty, $d$, and user ability, $u$: 
\begin{eqnarray*}
g(\pi_i) &=& \eta_i = \mu + \tau_{j(i)} +\delta_{k(i)}+ \nu_{s(i)} + u_{u(i)} + d_{d(i)},\\
Y &=& g^{-1}(\eta) + \varepsilon
\end{eqnarray*}
where $g$ is the logit link function, and $j(.), k(.), s(.), u(.)$, and $d(.)$ are  indexing functions that relate evaluation $i$ to the corresponding levels in the factor variables, to the observer, or a particular data sample. More specifically, $j(i) \in \{$Control, Standard, De-trended$\}$; $k(i) \in \{2,5,10\}$; $s(i) \in \{20, 30, 50, 75\}$; $u(i)$ maps to the participant's id of the $i$th evaluation; and $d(i)$ identifies the particular data sample used. 
Both user ability, $u$, and sample difficulty, $d$, are modeled as independent, normally distributed  random effects, i.e. $u_{u(i)} \sim N(0, \sigma_u^2)$, $d_{d(i)} \sim N(0,\sigma_d^2)$ with cov$(u, d) = 0$. We further assume that $E[\varepsilon] = 0$ and Var$[\varepsilon]=\sigma^2$.


\begin{table}[ht]
\centering
\caption{\label{tab:model} Coefficients and significances corresponding to  model $M_1$. The type of design is important for the power of a lineup. De-trended Q-Q plots lose a significant amount of power compared to both the regular and the standard version of Q-Q plots. }
\begin{tabular}{rrrrrl}
  \hline
 &\bf Estimate &\bf Std. Error &\bf z value &\bf Pr($>$$|$z$|$) & \\ 
  \hline
  Intercept &  -5.37 & 0.769 & -6.98 & 0.0000  & *** \\ [3pt]
\multicolumn{3}{l}{\bf design} \\
   Control & 0.00 & ----- & ----- & ----- \\ 
   Standard & 0.06 & 0.103 & 0.62 & 0.5371 \\
   De-trended & -0.50 & 0.104 & -4.77 & 0.0000 & ***\\  [3pt]
\multicolumn{4}{l}{\bf degrees of freedom} \\
  2 & 6.63 & 0.752 & 8.82 & 0.0000 & ***\\ 
  5 & 2.65 & 0.732 & 3.61 & 0.0003 & **\\ 
  10 & 0.00 & ----- & ----- & ----- \\ [3pt]
\multicolumn{3}{l}{\bf sample size} \\
  20 & 0.00 & ----- & ----- & ----- \\ 
  30 & 0.88 & 0.848 & 1.03 & 0.3014 \\ 
  50  & 3.26 & 0.837 & 3.90 & 0.0001 & ***\\ 
  75 & 2.20 & 0.838 & 2.63 & 0.0086  & **\\ 
   \hline
\multicolumn{6}{l}{Signif. codes:  0 $\le$ *** $\le$ 0.001 $\le$ ** $\le$ 0.01 $\le$ * $\le$ 0.05 $\le$ . $\le$ 0.1 $\le$ ' ' $\le$ 1}
\end{tabular}
\end{table}

The estimated model coefficients for model $M_1$ are shown in Table~\ref{tab:model}. 
Estimates of the variance components are $\widehat{\sigma}_u = 0.44$, $\widehat{\sigma}_d=1.95$, and $\widehat{\sigma} = 0.31$. Variances of user ability and data difficulty are large relative to residual variance, indicating that both random effects are necessary.
%
Compared to the difficulty level of lineups, participants' abilities only vary little. The difference between best and worst performance by participants has an effect of at most an estimated 
1.9-fold probability of detecting the data plot from a lineup. 

\section{Power: three different designs of Q-Q plots}\label{sec:power1}

%
As expected, the task of identifying non-normality becomes easier with increased sample size and more pronounced deviations from normality due to lower degrees of freedom. The  design of the Q-Q plot is of huge importance for the probability of choosing the data plot: compared to the control chart, add-on confidence bands help with evaluation in the standard design, but the difference is not significant.  Surprisingly, the de-trended Q-Q plot is significantly less powerful in detecting non-normality than either of the other designs. 
In terms of rejections of the null hypothesis, this means that normality is rejected in 24 out of the 48 lineups of the de-trended Q-Q plot. All of these cases are being rejected in all of the other designs as well, but using the control design another four lineups reject normality, and the standard design rejects yet another lineup.

\begin{figure}[hbt]
\begin{subfigure}{0.5\textwidth}
\begin{knitrout}
\definecolor{shadecolor}{rgb}{0.969, 0.969, 0.969}\color{fgcolor}
\includegraphics[width=\maxwidth]{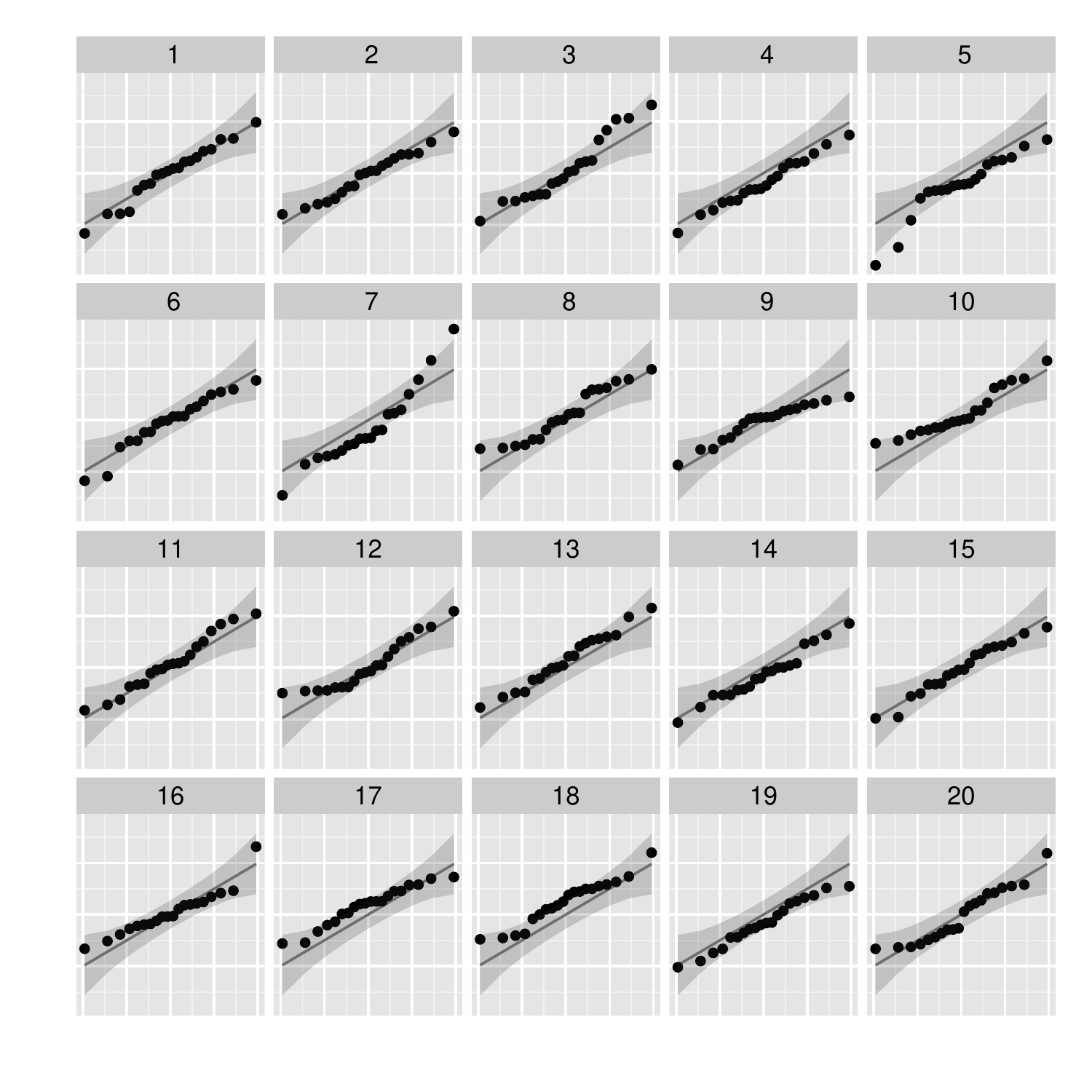} 

\end{knitrout}
\end{subfigure}
\begin{subfigure}{0.5\textwidth}
\begin{knitrout}
\definecolor{shadecolor}{rgb}{0.969, 0.969, 0.969}\color{fgcolor}
\includegraphics[width=\maxwidth]{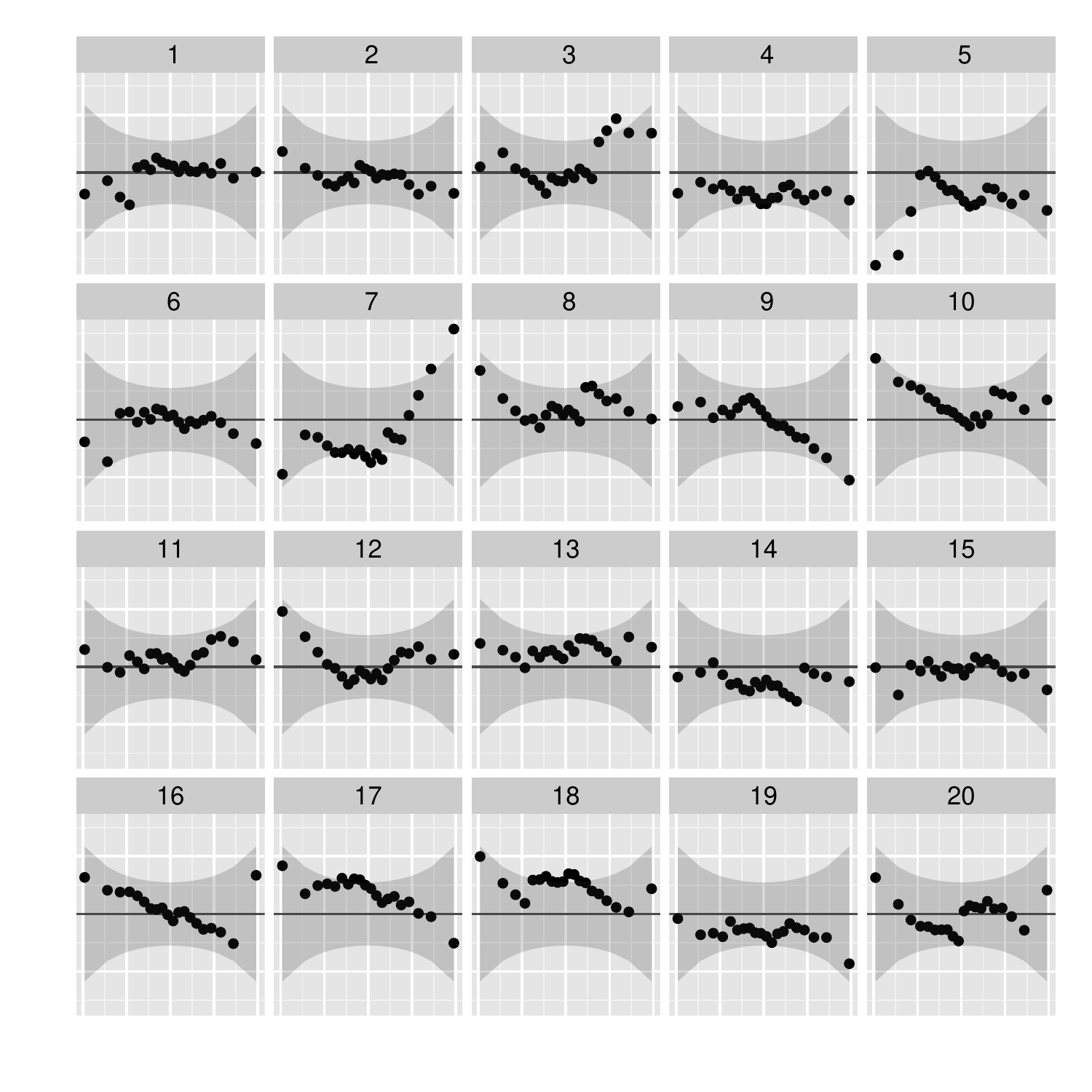} 

\end{knitrout}
\end{subfigure}
\caption{\label{fig:rotstd}Lineups of the same data in two different designs. In the standard design the data sample (in panel \#$(2^2+1)$) is identified 16 out of 27 times, leading to a rejection of normality. The same data set is only identified once out of 21 times in the lineup showing the de-trended design. Observers instead pick panel \#$(3^2-2)$ 18 times. }
\end{figure}

\begin{figure}[hbt]
\centering
\begin{knitrout}
\definecolor{shadecolor}{rgb}{0.969, 0.969, 0.969}\color{fgcolor}
\includegraphics[width=.65\textwidth]{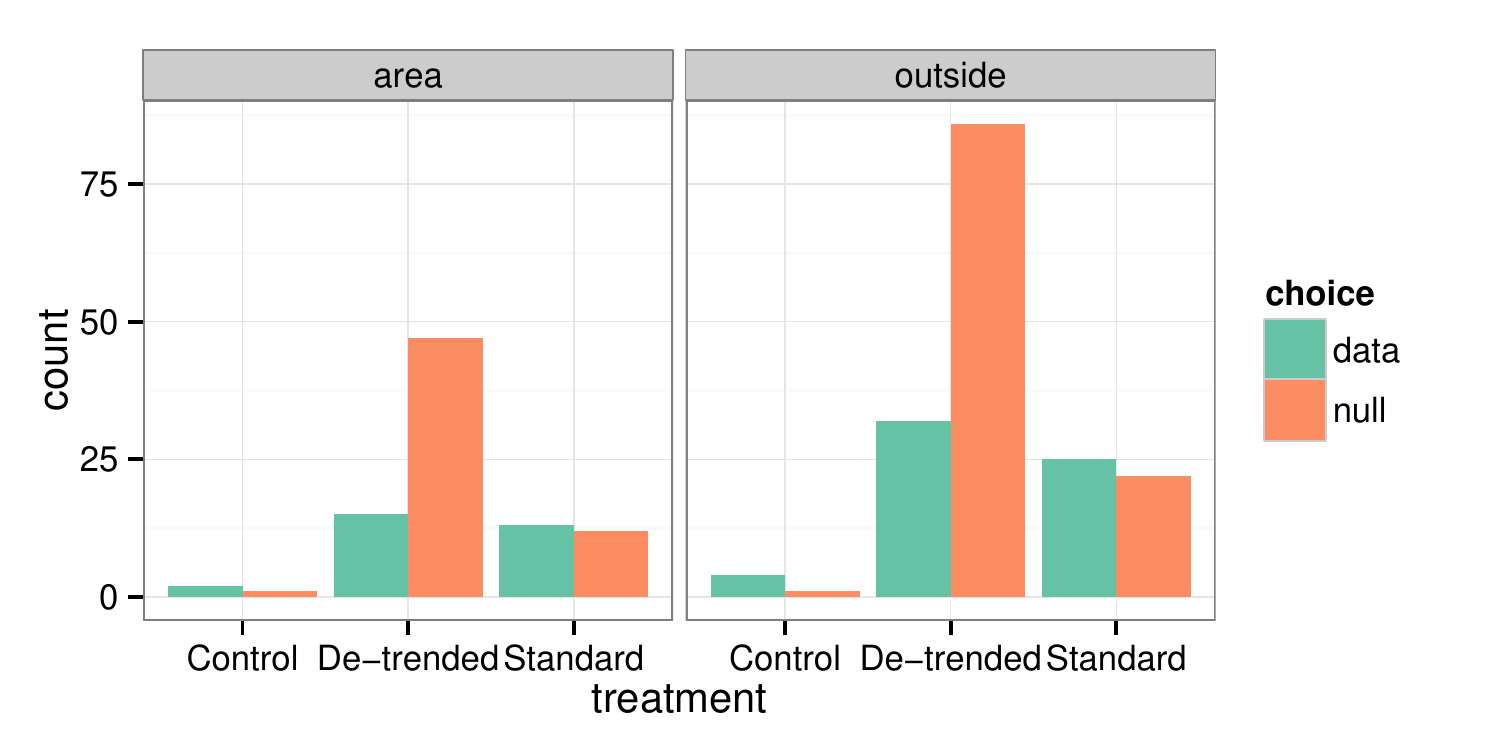} 

\end{knitrout}
\caption{\label{fig:rotfalse}Mentioning ``outside'' or ``area'' in the reason for selecting the plot from the lineup increases the probability of not identifying the data plot by a large factor in de-trended Q-Q plots. }
\end{figure}

%
To further investigate the difference between the standard and the de-trended designs, consider figure \ref{fig:rotstd}. Here, we have an example of a sample that is rejected based on a  lineup in the standard design, but not from a lineup of de-trended Q-Q plots. Instead of focusing on the panel showing the sample, observers focus on panel \#$(3^2-2)$ (with 18 out of 21 picks). This panel was picked as being most different 9 out of 27 times in the standard design, too, indicating that there is something special about it, but most observers (16 out of 27) picked the data in panel \#$(2^2+1)$ from the standard design. 
Two of the observers picking panel \#7 from the de-trended lineup thought that this panel was the one with the ``most dots outside the shaded area,'' i.e.~they focused on the middle of plot \#7. This is not a singular occurrence---when investigating  overall reasoning we take a closer look at the effect of the  words ``area'' and ``outside'' in the reason participants gave for making their choice of plot (figure~\ref{fig:rotfalse}). As expected, these words barely occur in the control design (where there is no shaded area). It seems to help in identifying the data plot in the standard design, but it severely  increases the chance of picking a null plot in the de-trended design. Why is that? The de-trended version is making better use of the space in the plot, it therefore  emphasizes deviations of points from the $x$-axis (i.e.~the theoretical distribution) and with it  the fact whether individual points are inside or outside the shaded areas corresponding to the (pointwise) 95\% confidence intervals. The responses suggest that participants take the shading very seriously and make their choice dependent on it. It also seems that the confidence bands mislead people---this suggests, that for the de-trended Q-Q plot design we might have to re-think how to display confidence intervals: it might be better to  either use  a more conservative  confidence level or change the approach altogether from pointwise confidence intervals to simultaneous confidence bands as, for example, discussed by \citet{Rosenkrantz:2000fd}.

Another promising approach might be to base confidence bands on the TS test \citep{buja:2013}. These bands define a test of normality and are narrower in the tails than those associated with the Kolmogorov-Smirnov test, while slightly more conservative in the middle of the distribution. These findings coincide with our observation that points  outside the confidence intervals were misleading participants, if this occurred in the middle of the plot.

\begin{figure}[hbt]
\centering
\begin{knitrout}
\definecolor{shadecolor}{rgb}{0.969, 0.969, 0.969}\color{fgcolor}
\includegraphics[width=\textwidth]{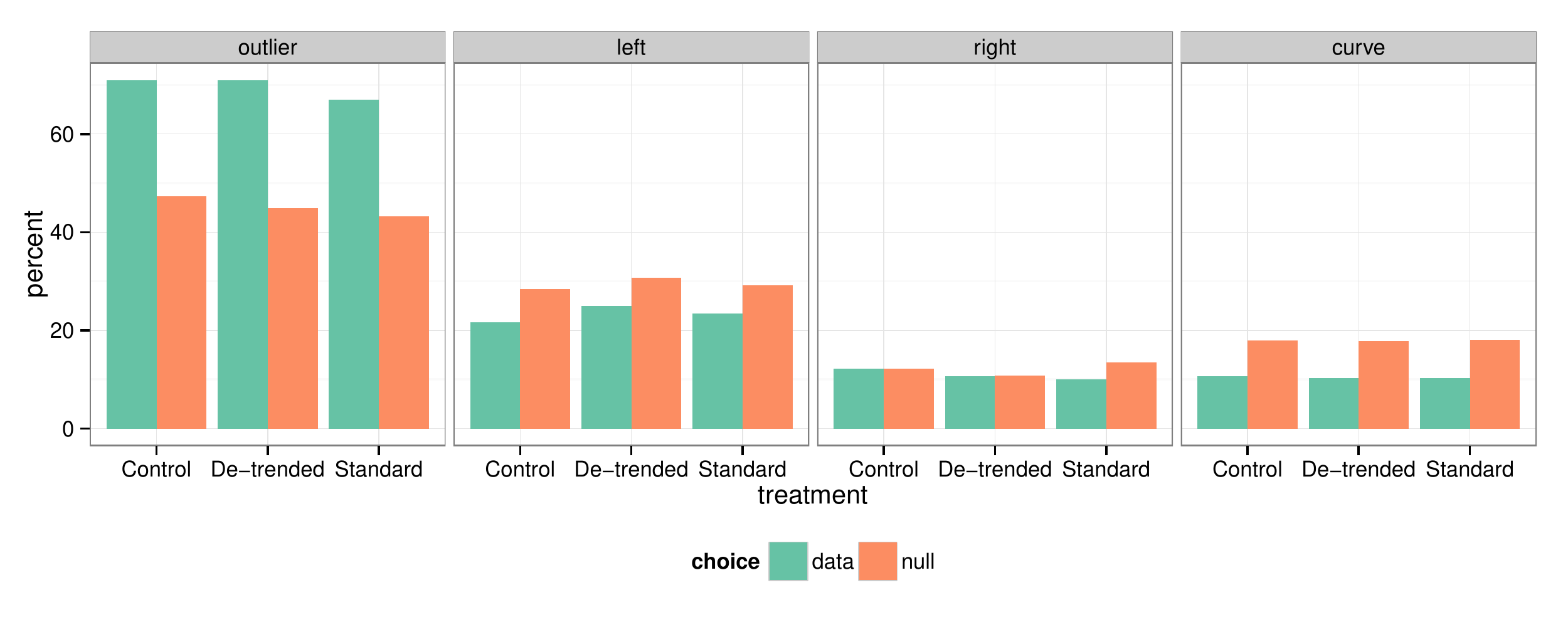} 

\end{knitrout}
\caption{\label{fig:choices}Overview of reasons participants gave for  their answer. ``Outliers'' as a reason drastically improves the chance of identifying the data plot. All the other reasons either have no effect or decrease the chance of picking the data. It is curious to see that so many more observers respond ``left side different'' over ``right;''  all the samples  come from a  $t$-distribution, so deviations in the extremes should therefore also be symmetric.}
\end{figure}

Figure~\ref{fig:choices} summarizes the reasons participants gave for the choice of plot they selected. Bars on the left show  identifications of the data plot, bars on the right represent selections of a null plot. The reasons represent the four reasons offered  to participants in check boxes. Notably, the reasons do not seem to differentiate between the three designs. Stating ``outliers'' as a reason for choosing a plot is helpful across all designs in picking the data out of the lineup. Stating ``left side different'' or ``points curve''  as the reason for choosing a plot decreases the chance for this plot to be the data plot. ``Right side different'' does not seem to have any effect. Interestingly, there is a big difference in the percentages of ``right'' and ``left.'' Participants favored to give ``left side different'' as a reason over ``right side different'', even though all distributions involved were symmetric, and therefore deviations from normality should also manifest themselves in a symmetric fashion.


\section{Power: visual and classical}\label{sec:power2}

It is important to recall that none of the data plots in the lineups were actually created using data from a normal distribution. Ideally, this should lead to rejection of the null hypothesis in every single instance.
This is not quite true, as can be seen in Table~\ref{tab:reject}, but what becomes evident is the high power  of visual inference. Based on lineups we are able to reject non-normality much more often than with any of the classical tests.

\begin{table}[ht]
\centering
\caption{\label{tab:reject}
 From left to right, we see the number of rejections from visual inference as well as the  Shapiro-Wilk, Anderson-Darling, Lilliefors,  Kolmogorov-Smirnov, and Cram\'er-von Mises tests for normality. Out of the 24 non-normal samples, 12 get rejected at the 5\% significance level based on evaluation by observers. None of the standard normal tests come close to that rejection rate. The power we observe here matches the power discussion by \citet{razali:2011} for the SW, AD, and the LF  test. The number in parentheses is the number of situations in which the Standard Q-Q plot agrees with the normal test in rejecting normality of the sample.}

\begin{tabular}{rrrrrrrrr}
  \hline
 & Standard  & SW & AD & LF   & CVM & \\ 
  \hline
  \hline
  reject $N(0,S^2)$  & 12  &  8 (7) & 5 (5) &  5 (5) & 4 (4) & \\ 
\hline
\end{tabular}
\end{table}

\begin{figure}
\centering
\includegraphics[width=0.5\textwidth]{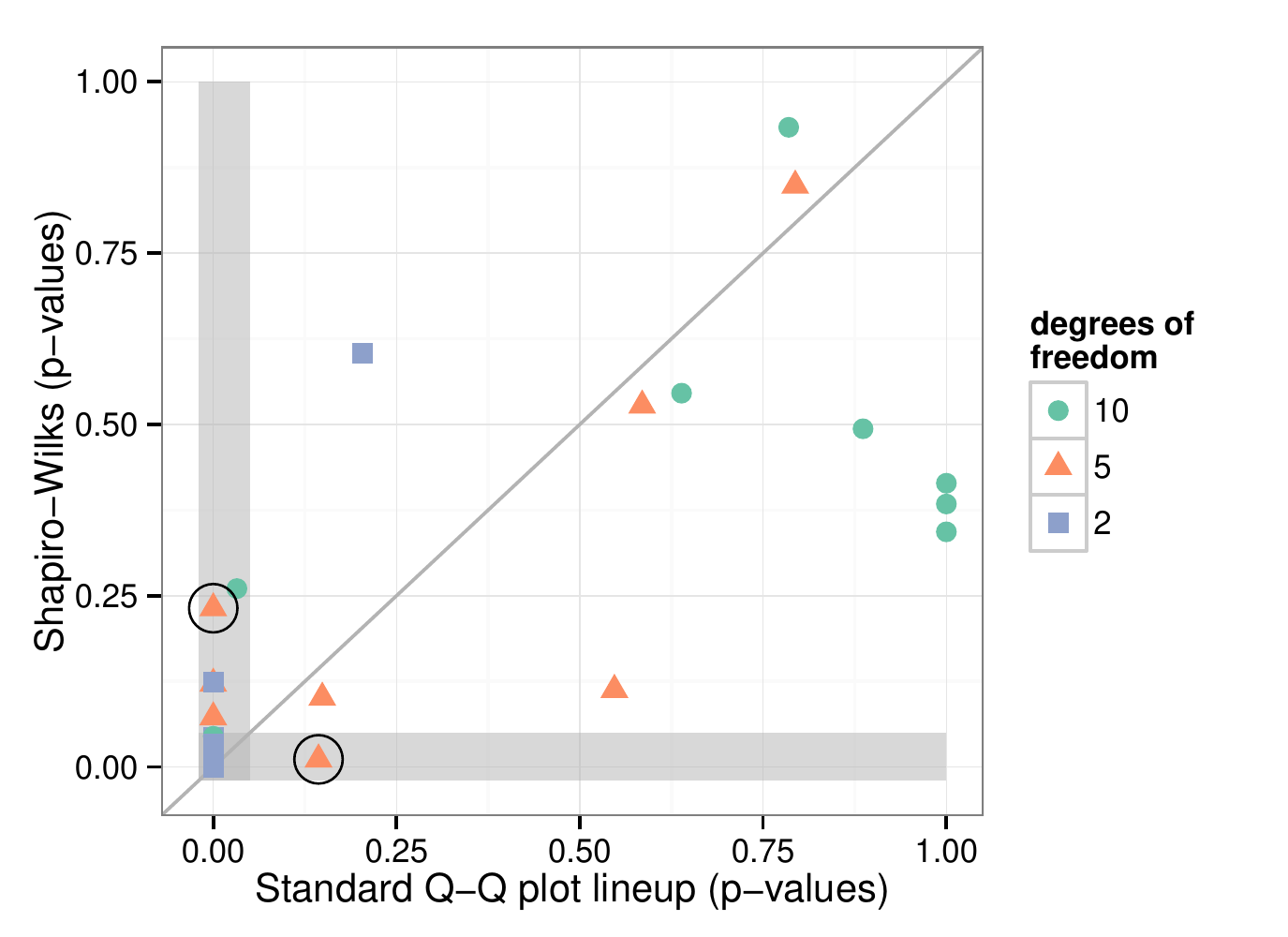} 
\caption{\label{fig:visnorm}  Scatterplot of $p$-values from the Shapiro-Wilk test and estimated $p$-values from lineups of Standard Q-Q plots. The grey shaded areas represent areas of rejection under at least one of the tests. The circled observations correspond to samples that lead to decidedly different decisions under the two tests. The lineups corresponding to these observations are shown in figure~\ref{fig:lpnorm}.}
\end{figure}

\begin{figure}
\begin{subfigure}[t]{.5\textwidth}

\includegraphics[width=\maxwidth]{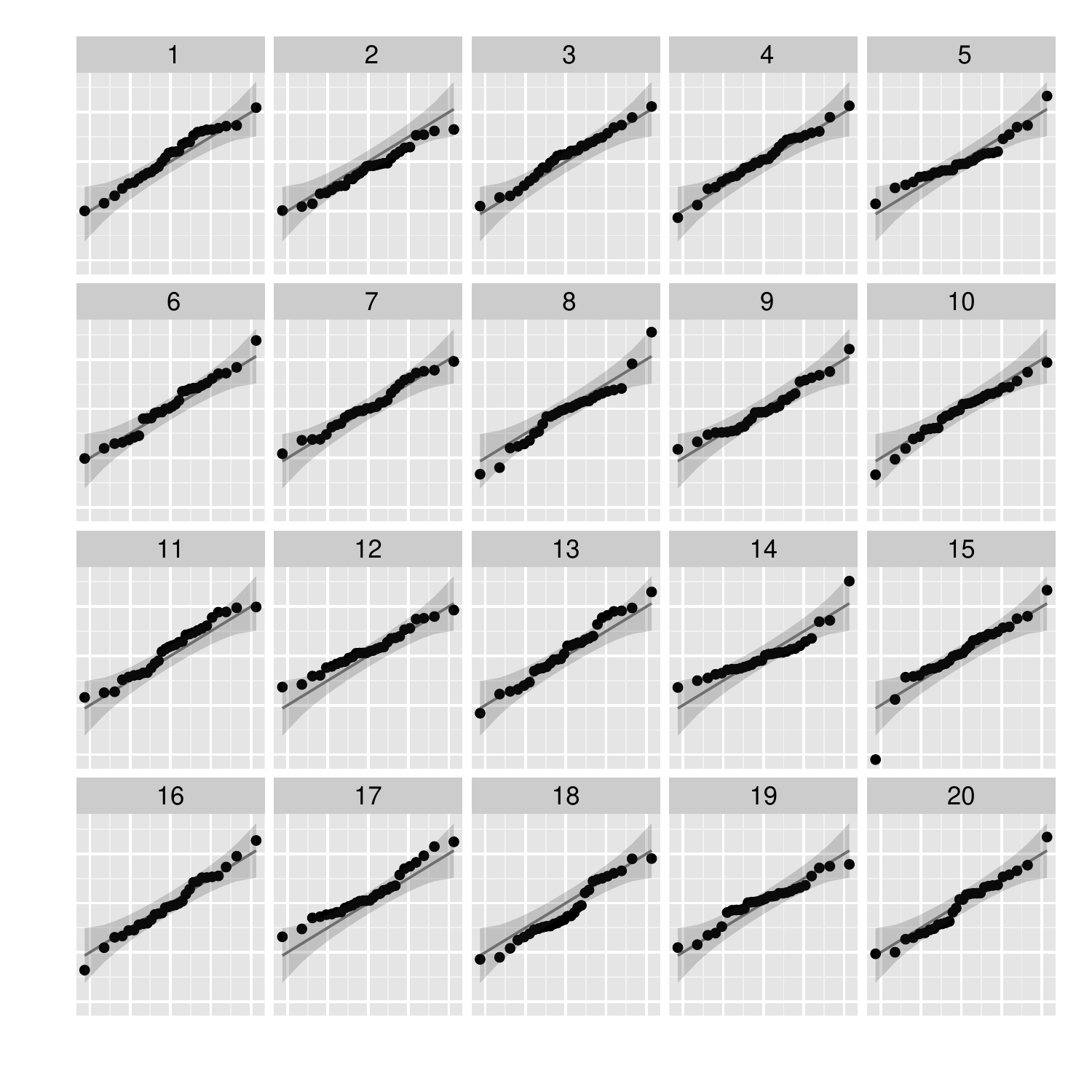} 

\hfill
\begin{tabular}{C{0.75cm}|C{0.75cm}|C{0.75cm}|C{0.75cm}|c}
   &   & 1 & 1 & 2 \\ 
   \hline
1.7 & 2 & 9.3 & 1 &   \\ 
   \hline
0.33 &   & 1.3 & 7 & \bf \it 1.5 \\ 
   \hline
  & 2 & 4 & 0.5 & 3.3 \\ 
  
\end{tabular}
\hfill

\end{subfigure}
\begin{subfigure}[t]{.5\textwidth}

\includegraphics[width=\maxwidth]{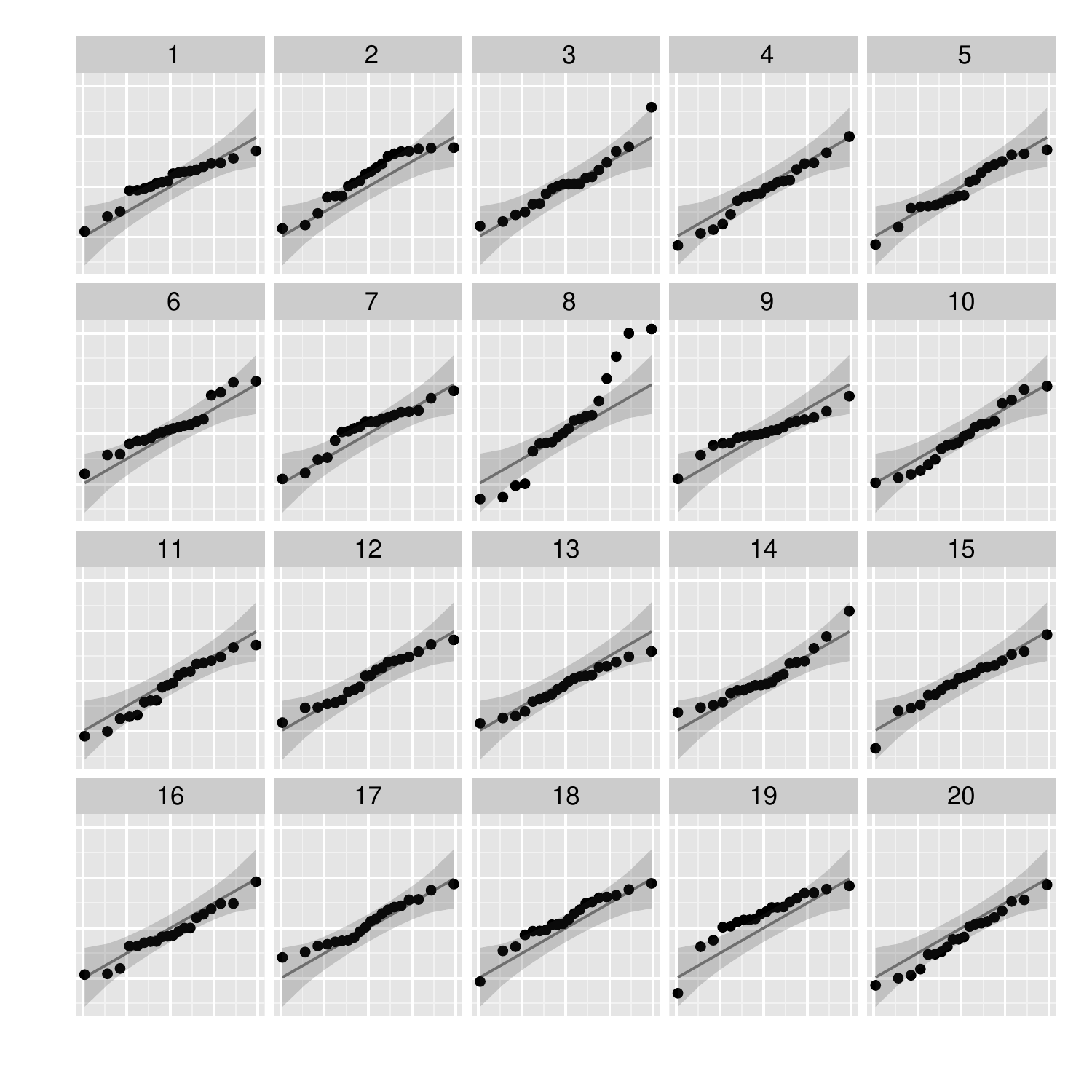} 

\hfill
\begin{tabular}{C{0.75cm}|C{0.75cm}|C{0.75cm}|C{0.75cm}|c}
   &   &   &   &   \\ 
   \hline
1 &   & \bf \it 23 &   &   \\ 
   \hline
0.5 &   &   &   &   \\ 
   \hline
  &   &   & 1 & 0.5 \\ 
  
\end{tabular}
\hfill

\end{subfigure}
\caption{\label{fig:lpnorm}  On the left, results are not significant, on the right they are highly significant. These two lineups correspond to the results circled in~figure~\ref{fig:visnorm}. The tables below the lineups show the number of times each of the panels was picked as the most different. Non-integer numbers result from multiple choice plots. The italicized numbers refer to the panel that contains the actual sample.
}
\end{figure}

Figure~\ref{fig:visnorm} shows a scatterplot of $p$-values from the SW test and estimated $p$-values from the lineup of Q-Q plots in the standard design. Out of the 24 samples, the tests agree on 18, of which seven are rejections. Of the remaining six, five are rejected only by the visual test, and one is rejected by SW, but not by the visual test. Two samples on which the tests disagree are circled in figure~\ref{fig:visnorm}. The two lineups corresponding to these observations are shown in figure~\ref{fig:lpnorm}. The lineup on the left corresponds to a sample that is rejected by the SW test, but is not rejected by the visual test: only 1.5 decisions (at least one observer picked two panels in his/her response) out of 38 identified the data panel as the most different, which is not enough to reject the null hypothesis of $N(0, S^2)$. 
In contrast, the data plot in the  lineup on the right is picked by 23 out of 26 independent observers, leading to a very clear rejection. 
The corresponding $p$-value in the SW test is 0.2318, after LF (0.1689) the test with the lowest $p$-value on this data sample out of all the normality tests.

\begin{figure}
\centering
\begin{knitrout}
\definecolor{shadecolor}{rgb}{0.969, 0.969, 0.969}\color{fgcolor}
\includegraphics[width=0.5\textwidth]{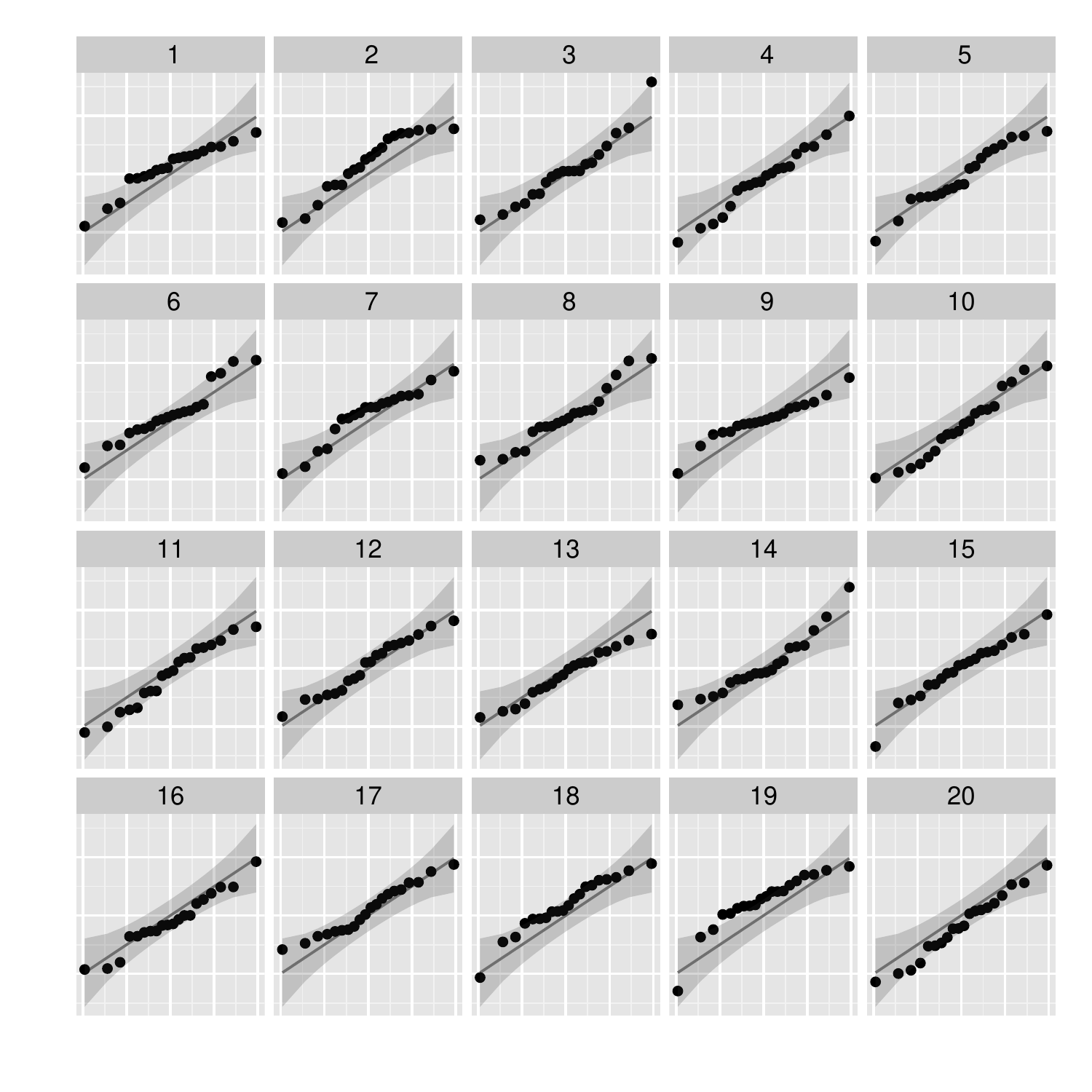} 

\end{knitrout}
\caption{\label{fig:lp3} Lineup of standard design Q-Q plots showing the data of the lineup on the right in figure~\ref{fig:lpnorm}. The hypothesized distribution is $N(0, \widehat{\sigma}^2)$, where $\widehat{\sigma}^2$ is estimated as the regular (i.e.~non-robust) sample variance, and nulls are drawn from that distribution. While not actually user tested, we do not think that the data stands out from the lineup.}
\end{figure}

The difference in significance between normality tests and the visual test might be due to the way the theoretical distribution against which the sample is compared is chosen. The normality tests are based on the sample mean and sample variance. Both of these estimates are affected by outliers. Compared to a normal distribution, the samples from a $t$ distribution exhibit heavier tails. In a finite sample, the heavier tails might look like outliers. By taking these outliers into account, the normality tests lose substantial power. The Q-Q plots, on the other hand, are based on a robust estimate of the scale based on the middle half of the empirical distribution. Q-Q plots are, therefore, less affected by outliers and the tails of a $t$ distribution are more easily distinguishable from the tails of a normal distribution, as can be seen in the lineup on the right of figure~\ref{fig:lpnorm}. Compare this to the lineup of figure~\ref{fig:lp3}, which is based on the same data, but the nulls are sampled from a normal distribution with a variance estimated as the sample variance. The data plot does not stand out, so we would not reject the null hypothesis based on this lineup. 
An inferior performance of normality tests based on sample mean and variance is also observed by \cite{buja:2013} in the discussion of the TS test. 

\section{Discussion and Conclusion}\label{sec:discussion}
In the comparison of the three designs of Q-Q plots, de-trended Q-Q plots turn out to have significantly less power in detecting non-normality than Q-Q plots in the standard and the control design. This is surprising, as results from cognitive psychology suggest that the de-trended version has superior qualities. From the additional reasoning provided by participants regarding their choice of plot it becomes obvious that this choice is mainly driven by points outside the shaded area depicting 95\% confidence intervals. This happens primarily in the middle of the distribution, confirming results by \citet{buja:2013}, and reopens the question of whether the design or the choice of confidence calculation is the reason for the inferiority. It would also make sense to fix the aspect ratio of plots in the de-trended design to make comparisons between the range of points in both axis directions possible.
All versions of Q-Q plots under consideration here are significantly better at detecting deviations from normality than classical normality tests. A contributing factor to this superior power might be  that in  Q-Q plots the whole sample is assessed rather than being reduced to the single value considered for the test statistic. 
Contributing to the power is also the robust estimation of the parameters for the normal distribution drawn as a line of fit in  Q-Q plots, while most normality tests are based on the outlier sensitive sample variance. 
This is consistent with findings in \citet{buja:2013} and  also poses the question of whether the power of classical normality tests might  be improved by using robust estimates for the mean and variance of the sample.

Q-Q plots are not restricted to the assessment of normality. In fact, they provide a general framework for testing any distributional assumptions. Used in the setting of lineups, they in particular allow an assessment of limiting distributions, i.e.~for example, lineups allow us to investigate distributions of
 samples from approximate normal or asymptotic normal distributions, for which  there exists  no classical test for finite sample sizes, whereas the lineup protocol provides us with a valid testing system
  as long as there is a method  to generate data under the null hypothesis for creating null plots in the lineup.

One of the drawbacks of the lineup test framework is that it comes at a higher cost, both monetary and in time, than classical testing.
However, developments such as {\tt nullabor} \citep{nullabor} or {\tt vis.test} \citep{vistest} allow us to be, at least to a degree, our own testers. For tests that are of a more sensitive nature, the cost of a test using a crowd-sourcing service is certainly a small enough item in the overall project budget that it is a feasible 
option. It also discourages the analyst from multiple testing!

Several possibilities for immediate extensions are obvious: the simulation study here is only concerned with deviations from normality as given by the $t$-distribution. Other types of deviations, such as skewed distributions or mixture distributions, would be interesting to consider as well. We doubt that the overall results would change dramatically, but it might provide more insight into what observers consider in making their assessments. 
The application of the lineup framework based on the related probability (P-P) plots poses a natural next question: \citet{koehler91} comment on the higher sensitivity of  P-P plots  to discrepancies in the middle of the distribution, such as caused by multiple modes. We can verify this and other statements for large samples and based on distributions. Lineups allow us to quantify the extent to which these statements hold for small sample sizes.

\bibliographystyle{asa}
\bibliography{qqplots}

\begin{thebibliography}{28}
\newcommand{\enquote}[1]{``#1''}
\expandafter\ifx\csname natexlab\endcsname\relax\def\natexlab#1{#1}\fi

\bibitem[{Aldor-Noiman et~al.(2013)Aldor-Noiman, Brown, Buja, Rolke, and
  Stine}]{buja:2013}
Aldor-Noiman, S., Brown, L.~D., Buja, A., Rolke, W., and Stine, R.~A. (2013),
  \enquote{The Power to See: A New Graphical Test of Normality,} \textit{The
  American Statistician}, 67, 249--260.

\bibitem[{Amazon(2010)}]{amazon}
Amazon (2010), \enquote{Mechanical {T}urk,}
  \url{https://www.mturk.com/mturk/welcome}.

\bibitem[{Anderson and Darling(1954)}]{adtest:1954}
Anderson, T.~W. and Darling, D.~A. (1954), \enquote{A Test of Goodness of Fit,}
  \textit{Journal of the American Statistical Association}, 49, 765--769.

\bibitem[{Becker et~al.(1988)Becker, Chambers, and Wilks}]{becker:s}
Becker, R.~A., Chambers, J.~M., and Wilks, A.~R. (1988), \textit{The New S
  Language: A Programming Environment for Data Analysis and Graphics},
  Monterey, CA, USA: Wadsworth and Brooks/Cole Advanced Books \& Software.

\bibitem[{Buja et~al.(2009)Buja, Cook, Hofmann, Lawrence, Lee, Swayne, and
  Wickham}]{buja:2009hp}
Buja, A., Cook, D., Hofmann, H., Lawrence, M., Lee, E.~K., Swayne, D.~F., and
  Wickham, H. (2009), \enquote{Statistical inference for exploratory data
  analysis and model diagnostics,} \textit{Philosophical Transactions of the
  Royal Society A: Mathematical, Physical and Engineering Sciences}, 367,
  4361--4383.

\bibitem[{Cleveland and McGill(1984)}]{cleveland:1984}
Cleveland, W.~S. and McGill, R. (1984), \enquote{Graphical Perception: Theory,
  Experimentation, and Application to the Development of Graphical Methods,}
  \textit{Journal of the American Statistical Association}, 79, 531--554.

\bibitem[{Cram\'{e}r(1928)}]{cramer:1928}
Cram\'{e}r, H. (1928), \enquote{On the composition of elementary errors,}
  \textit{Scandinavian Actuarial Journal}.

\bibitem[{Davison and Hinkley(1997)}]{Davison:1997}
Davison, A.~C. and Hinkley, D.~V. (1997), \textit{Bootstrap methods and their
  application}, Cambridge: Cambridge University Press.

\bibitem[{Gan et~al.(1991)Gan, Koehler, and Thompson}]{koehler91}
Gan, F.~F., Koehler, K.~J., and Thompson, J.~C. (1991), \enquote{Probability
  Plots and Distribution Curves for Assessing the Fit of Probability Models,}
  \textit{The American Statistician}, 45, 14--21.

\bibitem[{Hofmann et~al.(2012)Hofmann, Follett, Majumder, and
  Cook}]{Hofmann:2012ts}
Hofmann, H., Follett, L., Majumder, M., and Cook, D. (2012), \enquote{Graphical
  Tests for Power Comparison of Competing Designs,} \textit{Visualization and
  Computer Graphics, IEEE Transactions on}, 18, 2441--2448.

\bibitem[{Kolmogorov(1933)}]{kolmogorov:1933}
Kolmogorov, A.~N. (1933), \enquote{Sulla Determinazione Empirica di una Legge
  di Distribuzione,} \textit{Giornale dell'Istituto Italiano degli Attuari}, 4,
  83--91.

\bibitem[{Lilliefors(1967)}]{lilliefors}
Lilliefors, H. (1967), \enquote{On the Kolmogorov–Smirnov test for normality
  with mean and variance unknown,} \textit{Journal of the American Statistical
  Association}, 62, 399--402.

\bibitem[{Majumder et~al.(2013)Majumder, Hofmann, and Cook}]{mahbub:2013}
Majumder, M., Hofmann, H., and Cook, D. (2013), \enquote{Validation of Visual
  Statistical Inference, Applied to Linear Models,} \textit{Journal of the
  American Statistical Association}, 108, 942--956.

\bibitem[{Razali and Wah(2011)}]{razali:2011}
Razali, N. and Wah, Y.~B. (2011), \enquote{{Power comparisons of Shapiro-Wilk,
  Kolmogorov-Smirnov, Lilliefors and Anderson-Darling tests},} \textit{Journal
  of Statistical Modeling and Analytics}, 2, 21--33.

\bibitem[{Robbins(2005)}]{robbins:2005}
Robbins, N. (2005), \textit{Creating More Effective Graphs}, Wiley.

\bibitem[{Rosenkrantz(2000)}]{Rosenkrantz:2000fd}
Rosenkrantz, W.~A. (2000), \enquote{{Confidence Bands for Quantile Functions: A
  Parametric and Graphic Alternative for Testing Goodness of Fit},} \textit{The
  American Statistician}, 54, 185--190.

\bibitem[{Rousseeuw and Croux(1993)}]{rousseeuw}
Rousseeuw, P.~J. and Croux, C. (1993), \enquote{Alternatives to the Median
  Absolute Deviation,} \textit{Journal of the American Statistical
  Association}, 88, 1273--1283.

\bibitem[{Shapiro and Wilk(1965)}]{Shapiro:1965kt}
Shapiro, S.~S. and Wilk, M.~B. (1965), \enquote{{An Analysis of Variance Test
  for Normality (Complete Samples)},} \textit{Biometrika}, 52, 591--611.

\bibitem[{Smirnov(1948)}]{smirnov:1948}
Smirnov, N. (1948), \enquote{Table for Estimating the Goodness of Fit of
  Empirical Distributions,} \textit{The Annals of Mathematical Statistics}, 19,
  279--281.

\bibitem[{Snow(2013)}]{vistest}
Snow, G. (2013), \textit{{TeachingDemos}: Demonstrations for teaching and
  learning}, {R} package version 2.9.

\bibitem[{Stephens(1974)}]{stephens:1974}
Stephens, M.~A. (1974), \enquote{{EDF} Statistics for Goodness of Fit and Some
  Comparisons,} \textit{Journal of the American Statistical Association}, 69,
  730--737.

\bibitem[{Thode(2002)}]{thode:2002}
Thode, H.~C. (2002), \textit{Testing for Normality}, New York: Marcel Dekker.

\bibitem[{Tukey(1977)}]{tukey:eda}
Tukey, J.~W. (1977), \textit{Exploratory Data Analysis}, Addison-Wesley, 1st
  ed.

\bibitem[{{Vander Plas} and Hofmann(2014)}]{sineillusion}
{Vander Plas}, S. and Hofmann, H. (2014), \enquote{Signs of the Sine Illusion
  -- and why we need to care,} \textit{Journal of Computational and Graphical
  Statistics}, to appear.

\bibitem[{{von Mises}(1928)}]{mises:1928}
{von Mises}, R.~E. (1928), \textit{Wahrscheinlichkeit, Statistik und Wahrheit},
  Julius Springer.

\bibitem[{Wainer(2000)}]{wainer:2000}
Wainer, H. (2000), \textit{Visual Revelations}, Psychology Press.

\bibitem[{Wickham(2012)}]{nullabor}
Wickham, H. (2012), \textit{{n}ullabor: Tools for graphical inference}, {R}
  package version 0.2.1.

\bibitem[{Wilk and Gnanadesikan(1968)}]{Wilk:1968}
Wilk, M.~B. and Gnanadesikan, R. (1968), \enquote{Probability Plotting Methods
  for the Analysis of Data,} \textit{Biometrika}, 55, 1--17.

\end{thebibliography}

\end{document}